\documentclass[pra,aps,superscriptaddress,twocolumn,floatfix]{revtex4-1}

\usepackage{dsfont}
\usepackage{mathtext}

\usepackage{mathtools}
\usepackage[usenames,dvipsnames]{xcolor}
\usepackage{amsmath}
\usepackage{amssymb,mathrsfs}
\usepackage{graphicx}
\usepackage{epstopdf}

\usepackage{mathtext}

\usepackage{bm}

\newcommand{\br}{{\bm r}}

\begin{document}
	
	\title{Edge Solitons in Lieb Topological Floquet Insulators}

	\author{S. K. Ivanov}
	\affiliation{Moscow Institute of Physics and Technology, Institutsky
		lane 9, Dolgoprudny, Moscow region, 141700, Russia}
	\affiliation{Institute of Spectroscopy, Russian Academy of Sciences, Fizicheskaya Str., 5, Troitsk, Moscow, 108840, Russia}
	
	\author{Y. V. Kartashov}
	\affiliation{Institute of Spectroscopy, Russian Academy of Sciences, Fizicheskaya Str., 5, Troitsk, Moscow, 108840, Russia}
	\affiliation{ICFO-Institut de Ciencies Fotoniques, The Barcelona Institute of Science and Technology, 08860 Castelldefels (Barcelona), Spain}
	
	\author{L. J. Maczewsky}
	\affiliation{Institute for Physics, University of Rostock, Albert-Einstein-Str. 23, 18059 Rostock, Germany}
	
	\author{A. Szameit}
	\affiliation{Institute for Physics, University of Rostock, Albert-Einstein-Str. 23, 18059 Rostock, Germany}
	
	\author{V. V. Konotop}
	\affiliation{Departamento de F\'isica, Faculdade de Ci\^encias,
		Universidade de Lisboa, Campo Grande, Ed. C8, Lisboa 1749-016, Portugal, and Centro de F\'isica Te\'orica e Computacional, Universidade de Lisboa, Campo Grande, Ed. C8, Lisboa 1749-016, Portugal}
	
	\begin{abstract}
		We describe topological edge solitons in a continuous dislocated Lieb array of helical waveguides. The linear Floquet spectrum of this structure is characterized by the presence of two topological gaps with edge states residing in them. A focusing nonlinearity enables families of topological edge solitons bifurcating from the linear edge states. Such solitons are localized both along and across the edge of the array. Due to the non-monotonic dependence of the propagation constant of the edge states on the Bloch momentum, one can construct topological edge solitons that either propagate in different directions along the same boundary or do not move. This allows us to study collisions of edge solitons moving in the opposite directions. Such solitons always interpenetrate each other without noticeable radiative losses; however, they exhibit a spatial shift that depends on the initial phase difference.
	\end{abstract}

\maketitle

Topological insulation~\cite{elect1,elect2} is a fundamental phenomenon that spans across several areas of physics. Topological photonics, initiated by the seminal paper~\cite{Haldane}, is one of such areas attracting nowadays growing attention~\cite{topphot1,topphot2}. Floquet insulators \cite{floquet01,floquet02} are a particular form of topological insulators, which are characterized by special topological invariants \cite{Rudner}. In such systems, a periodic modulation in the evolution coordinate breaks time-reversal symmetry, resulting in the appearance of the in-gap unidirectional edge states. Photonic Floquet topological insulators have been realized with helical waveguide arrays \cite{helix1} and were also explored in more complex modulated structures \cite{helix2,helix3,Leykam02}, including quasicrystals \cite{quasicrystal01}.

Topological effects in optical systems of helical waveguides can be combined with nonlinear self-action, enabling a plethora of phenomena including modulational instabilities~\cite{instabil,Leykam01}, and the existence of topological edge solitons. Edge solitons have been obtained numerically in continuous~\cite{Leykam01}, discrete ~\cite{AblMa2014,AblCole2017,AblCole2019}, and Dirac \cite{SSLK-19} models. They are essentially two-dimensional objects, propagating along the boundary of the topological insulator with the velocity imposed by the group velocity of the Floquet edge state on which soliton is constructed. In previously considered Floquet insulators such solitons, obtained for a given interface, were always co-propagating and suffering from considerable radiative losses due to strong localization~\cite{Leykam01}. In such setting it was practically impossible to implement interactions of edge solitons with appreciably different Bloch momenta of the carrier waves on finite distances in finite samples. 

In this Letter we show that it is possible to obtain edge solitons that move, along the same edge, in the opposite directions or even remain immobile in specially designed Floquet insulators with non-monotonous topological branches of the spectrum. We obtain them in different gaps of real-world continuous system - dislocated Lieb array of helical waveguides. Counter-propagating Floquet edge solitons interact almost elastically, but acquire phase-dependent spatial shift after collision.

\begin{figure}[t!]
\centering
\includegraphics[width=1\linewidth]{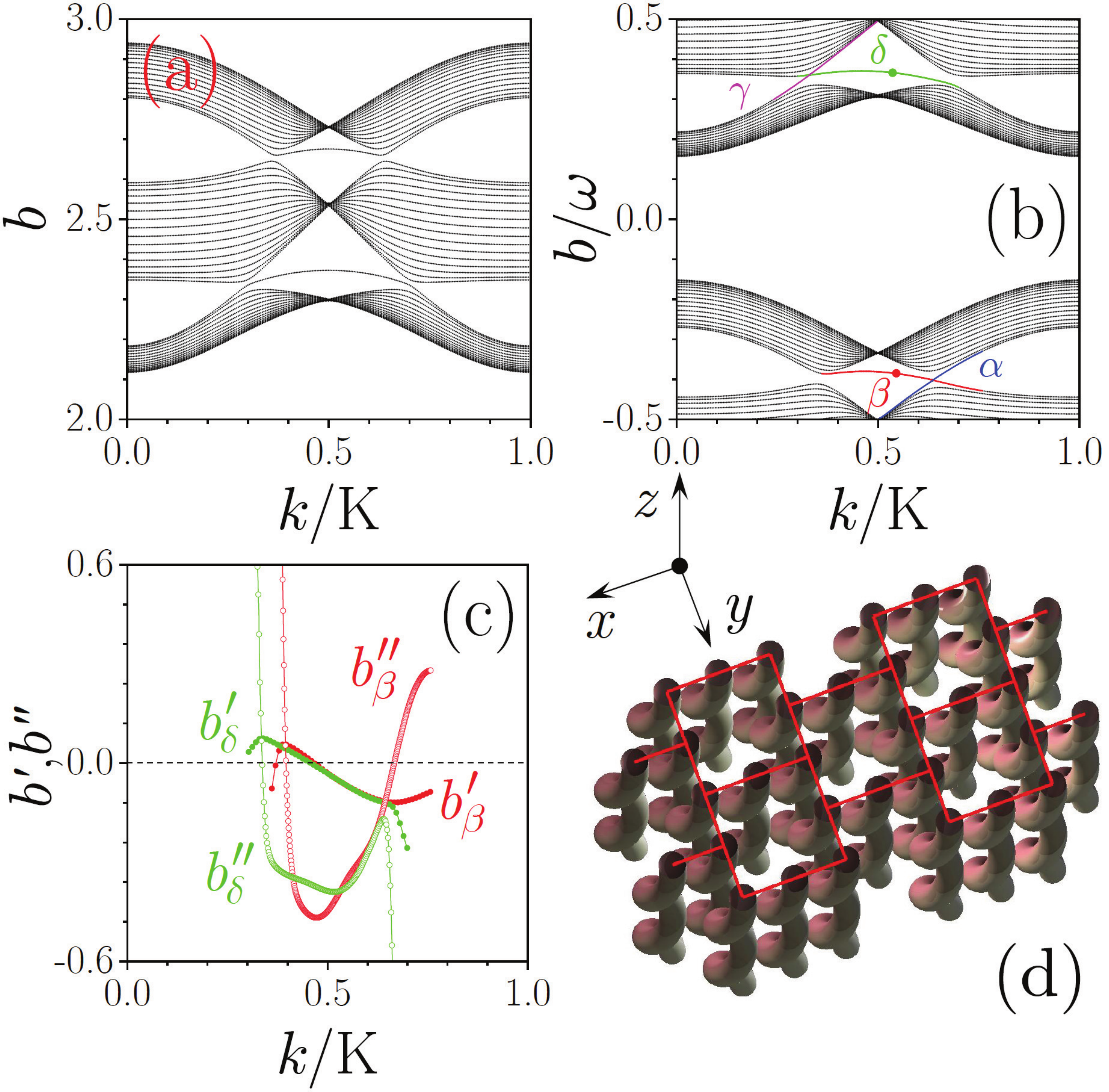}
\caption{(a) Propagation constants vs normalized Bloch momentum $k/K$ for a dislocated Lieb array with straight waveguides and (b) propagation constants vs $k/K$ for array with helical waveguides for helix period $Z=6$ and radius $r_0=0.4$. Dots indicate modes on which solitons from Fig.~\ref{fig:two} are built on. (c) Velocities $b^\prime$ (solid dots) and dispersion $b^{\prime\prime}$ (open dots) for the edge states associated with red and green topological branches in (b). (d) Schematic illustration of the dislocated Lieb array.}
\label{fig:one}
\end{figure}

\begin{figure*}[t!]
\centering
\includegraphics[width=\linewidth]{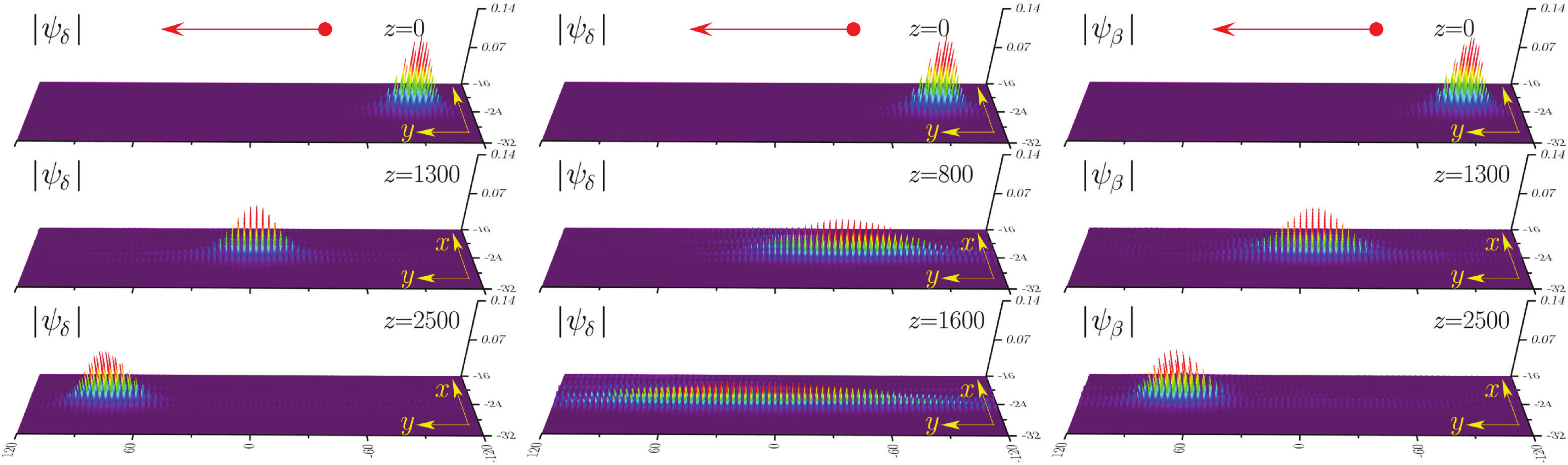}
\caption{(Left) Evolution of the edge soliton from green branch at $k =0.54 K$ corresponding to $b^{\rm nl}_{\delta} =0.004$, $b_\delta^\prime =-0.062$, $b_\delta^{\prime\prime} =-0.384$, $\chi_\delta =0.458$. (Center) Evolution of the same input, but in linear array. (Right) Evolution of the edge soliton from red branch at $k=0.53K$ corresponding to $b^{\rm nl}_{\beta} =0.004$, $b_\beta^\prime=-0.059$, $b_\beta^{\prime\prime}=-0.414$, $\chi_\beta =0.402$. Red arrows indicate direction of motion. The ratio of widths of soliton in the directions along and across the interface in the left and right columns is  $\sim5$.}
\label{fig:two}
\end{figure*}

We describe the propagation of a paraxial light beam along the $z$ axis of a helical waveguide array with focusing cubic (Kerr) nonlinearity using the nonlinear Schr\"{o}dinger (NLS) equation for the dimensionless field amplitude $\psi$: 
\begin{equation} 
\label{NLS_dimensionless}
    i \frac{\partial \psi}{\partial z} = -\frac{1}{2} \nabla^2_\perp \psi - |\psi|^2\psi - \mathcal{R}(\textbf{r},z) \psi.
\end{equation}
Here $\nabla_\perp=(\partial_x,\partial_y)$ and $\textbf{r}=x \textbf{e}_x+y \textbf{e}_y$ is the radius-vector in the transverse plane defined by the mutually  orthogonal unit vectors $\textbf{e}_x$ and $\textbf{e}_y$. The $Z$-periodic function $\mathcal{R}(\textbf{r},z)=p\sum_{n,m}\exp{[-((x'-x_n)^2+(y'-y_m)^2)^2/a^4]}$ with $x'=x-r_0\sin(\omega z)$ and $y'=y+r_0-r_0\cos(\omega z)$, where $\omega = 2\pi/Z$ the rotation frequency, describes the refractive index 
of the array of waveguides characterized by the depth $p$, width $a$, and the helix radius $r_0$. The centers of rotation of helical waveguides are placed in the nodes $(x_n,y_m)$ of the dislocated Lieb grid characterized by spacing $d$ between the neighbour nodes [see Fig.~\ref{fig:one}(d) for schematic illustration]. The array is infinite along the $y$-axis and truncated along the $x$-axis. We consider array truncation creating bearded interface on the left and straight interface on the right. Further we set $d=1.6$, $a=0.4$, and $p=10$ that corresponds to $16~\mu\textrm{m}$ separation between neighbouring waveguides of width $4~\mu\textrm{m}$, and refractive index modulation depth of $1.1\cdot 10^{-3}$ at the wavelength $\lambda=800~\textrm{nm}$, in accordance with experiments \cite{helix1}. Effective nonlinear refractive index is $n_2\sim 1.4\cdot 10^{-20}~\textrm{m}^2/\textrm{W}$. Typical helix radius $r_0=0.4$ ($4~\mu \textrm{m}$) and longitudinal period $Z=6$ ($6.8~\textrm{mm}$) guarantee low radiative losses in this structure.

Eigenmodes of the helical waveguide array with boundaries along the $y$-direction are the Floquet-Bloch waves $\psi_{\nu,k}(\textbf{r},z)=\phi_{\nu,k}(\textbf{r},z)\exp{(ib_{\nu,k}z)} =u_{\nu,k}(\textbf{r},z)\exp{(ib_{\nu,k}z+iky)}$ where $\phi_{\nu,k}(\textbf{r},z)$ are periodic along the $z$-axis, $\phi_{\nu,k}(\textbf{r},z)=\phi_{\nu,k}(\textbf{r},z+Z)$, while  $u_{\nu,k}(\textbf{r},z)$ are periodic along the $y$ and $z$ axes: $u_{\nu,k}(\textbf{r}+L\textbf{e}_y,z)=u_{\nu,k}(\textbf{r},z+Z)=u_{\nu,k}(\textbf{r},z)$ (notice that $y$-period is $L=2d$), $k$ is the Bloch momentum varying within transverse Brillouin zone of width $K=2\pi/L$, $\nu$ is either the band index or an index of the edge state (they can be ordered, say, by decreasing value of $b_{\nu,k}$ for a given $k$, where $b_{\nu,k}\in [-\omega/2,+\omega/2)$, defined modulo $\omega$~\cite{Moore,Rudner}, is a $k$-dependent eigenvalue of the problem~\cite{floquet01} $[i \partial_z +(1/2) \nabla^2_\perp + \mathcal{R}(\textbf{r},z)] \phi_{\nu,k} =b_{\nu,k}\phi_{\nu,k}$. Edge states are localized along the $x$ axis, i.e., $u_{\nu,k}|_{x\rightarrow\pm\infty}\rightarrow0$. 

The linear spectrum  of an infinite dislocated Lieb array with straight ($r_0=0$) waveguides was found using the standard plane-wave expansion method: it is shown in Fig.~\ref{fig:one}(a) for $k\in[0,K]$. {In this case $ \phi_{\nu,k}$ is $z$-independent and $b$ is a conventional propagation constant} (for illustrative purposes we omitted indices $\nu$ and $k$ in axis labels $b$ in Fig. 1).
Because the unit cell of the infinite dislocated Lieb array contains three waveguides, there are three bands in the top group of bands. The branches between bands in Fig.~\ref{fig:one}(a) arise due to truncation and correspond to the non-topological edge states. This picture changes qualitatively for helical waveguides with $r_0\neq0$  [Fig.~\ref{fig:one}(b)]. {Now $b_{\nu,k}$ can be viewed as a quasi-propagation constant of $Z$-averaged dynamics (it is the analog of the quasi-energies in quantum systems) and for brevity will be referred below as a propagation constant.}
Technically, the Floquet spectrum of the helical array was obtained using propagation-projection method~\cite{Leykam01,Leykam02}. Waveguide rotation leads to opening of the topological gaps around special points in the spectrum. The widths of the gaps increase with increasing $r_0$ and decreasing $Z$. If such an array is truncated, topological in-gap edge states branching off the boundaries of the bulk bands appear. In Fig.~\ref{fig:one}(b) we show the topological branches appearing at the right straight edge, highlighted with blue $(\alpha)$ and magenta $(\gamma)$ colours, and at the left bearded edge, highlighted with red $(\beta)$ and green $(\delta)$, with black curves corresponding to the bulk modes. 
The  dislocated Lieb array does not feature flat bands, unlike conventional Lieb arrays \cite{Lieb01,Lieb02,Lieb03}, but offers a number of advantages for existence of solitons. Namely, a usual Lieb array for the same (realistic) $r_0$ and $Z$ parameters does not allow for sufficiently large detunings of the propagation constant from linear topological edge levels, what is required for stable bright soliton formation. Additionally, topological levels of the dislocated array extend to the interval of momenta $k$ offering larger domain of soliton existence.

\begin{figure*}[t!]
\centering
\includegraphics[width=\linewidth]{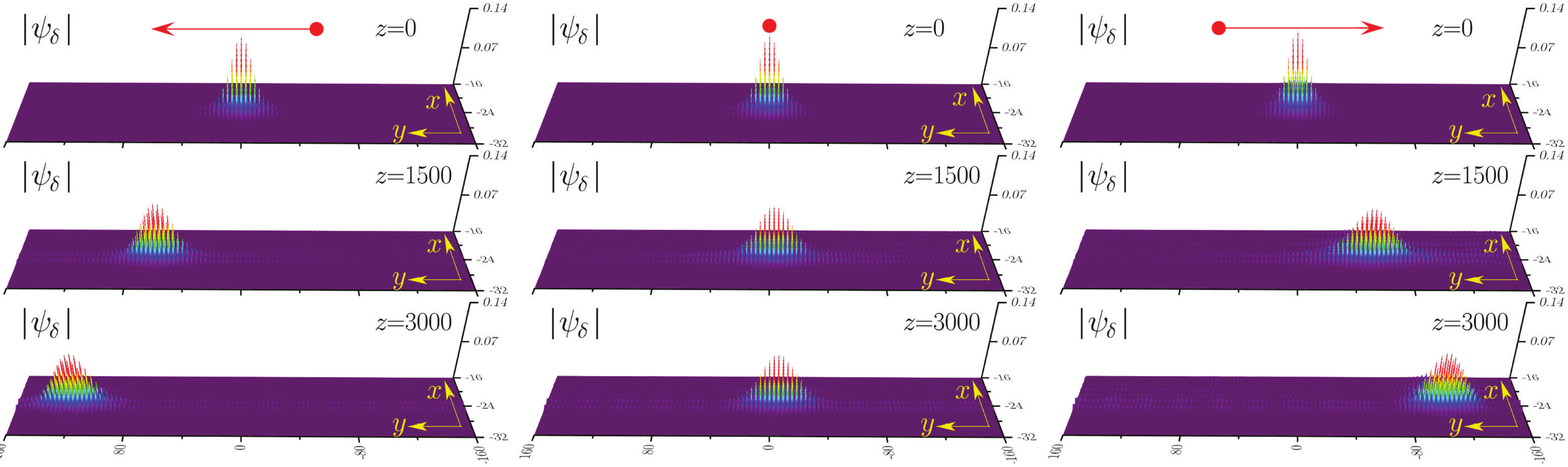}
\caption{(Left) Edge soliton moving in the positive $y$‐direction at $k = 0.51K$, $b_\delta^\prime=-0.039$, $b_\delta^{\prime\prime}=-0.392$, $\chi_\delta=0.488$, (center) standing edge soliton at $k = 0.46 K$, $b_\delta^\prime =0.0$, $b_\delta^{\prime\prime}=-0.373$, $\chi_\delta=0.462$, and (right) edge soliton moving in the negative $y$‐direction at $k = 0.42 K$, $b_\delta^{\prime}=0.028$, $b_\delta^{\prime\prime}=-0.356$, $\chi_\delta=0.377$. All states are constructed on the same green branch, in all cases $b^{\rm nl}_{\delta}=0.004$.}
\label{fig:three}
\end{figure*}

To consider bifurcation of the family of edge solitons from a linear topological state we use the slowly-varying-amplitude approximation and express nonlinear modes bifurcating from the branches $\nu$ (where $\nu=\alpha,\beta,\delta,\gamma$, see Fig.~\ref{fig:one} (b)) as $\psi_ {\nu}\approx e^{ib_{\nu,k} z}A_{\nu}(Y,z)\phi_{\nu, k}$, where $A_{\nu}$ is the slowly varying amplitude and $Y=y-v_{\nu,k}z$ is the coordinate in the frame moving with velocity $v_{\nu,k}=-b_{\nu}^{\prime}$  (hereafter we use notations $b_{\nu}^\prime=\partial b_{\nu,k}/\partial k$ and $b_{\nu}^{\prime\prime}=\partial^2 b_{\nu,k}/\partial k^2$, i.e., we do not indicate explicitly the Bloch momentum $k$ at which the family bifurcates). One can show that $A_{\nu}$ solves the NLS equation
\begin{eqnarray} \label{NLS_main}
    i\frac{\partial A_\nu}{\partial z} - \frac{b_{\nu}^{\prime\prime}}{2}\frac{\partial^2 A_\nu}{\partial Y^2} + \chi_\nu |A_\nu|^2A_\nu = 0,
\end{eqnarray}
with the nonlinear coefficient $
\chi_\nu=\langle (|\phi_{\nu,k}|^2,|\phi_{\nu,k}|^2)\rangle_Z$ defined using the inner product $(f,g)=\int_{S}f^*(\br,z) g(\br,z)d\br$, where the integral is computed over the whole area of the lattice $S$, as well as averaging over the helix period given by $\langle f\rangle_Z=Z^{-1}\int_{0}^{Z}f(\textbf{r},z)dz$. Here we used the fact that carrier Bloch waves can be always chosen orthogonal and normalized:
$
    (\phi_{\nu,k}(\br,z),\phi_{\nu',k'}(\br,z))=\delta_{\nu\nu'}
    \delta_{kk'}
$
($\delta_{ij}$ is the Kronecker delta). Eq.~(\ref{NLS_main}) predicts the formation of bright envelope solitons described by $A_{\nu}= (2b_{\nu}^{nl}/\chi_\nu)^{1/2} \; \mathrm{sech}(2^{1/2} Y/\ell_{\nu,k})e^{-ib_{\nu}^{nl}z}$, where $\ell_{\nu,k}=(-b''_{\nu}/b_{\nu}^{nl})^{1/2}$ is the soliton width. Since in focusing nonlinear medium $\chi_\nu>0$, one can expect soliton bifurcation from the Floquet-Bloch edge state $\psi_{\nu,k}$ if at a given momentum $k$ the effective-dispersion coefficient $b_{\nu}^{\prime\prime}$ is negative. The effective group velocity and dispersion for topological branches $\beta$ and $\delta$ from the left edge of the array are shown in Fig.~\ref{fig:one}(c). The bifurcation is parametrized by the nonlinear shift $b_\nu^{\rm nl}$ of the propagation constant from $b_{\nu,k}$ ($b_\nu^{\rm nl}\to 0$ corresponds to the linear limit). The shift $b_\nu^{\rm nl}$ must be small to ensure the validity of (\ref{NLS_main}), i.e., validity of the scaling $b_{\nu}^{\rm nl}A_\nu \sim \partial A_\nu/\partial z\sim \partial^2 A_\nu/\partial z^2\sim \chi_\nu |A_\nu|^2A_\nu$. 

Lieb lattices of helical waveguides feature two topological gaps (and one non-topological gap) in the Floquet spectrum. This enables co-existence of topological solitons from different gaps at the same edge. Such solitons bifurcate from different topological branches in the overlapping intervals of momenta $k$ [see $\beta$ and $\delta$ branches Fig.~\ref{fig:one} (b)]. Left and right columns of Fig.~\ref{fig:two} show propagation of such edge solitons bifurcating from points marked by dots in Fig.~\ref{fig:one}(b) in different gaps and constructed using envelope solution of Eq.~(\ref{NLS_main}). Here we chose $b^{\rm nl}_{\nu}$ small enough to ensure the validity of (\ref{NLS_main}), and at the same time for selected $b^{\rm nl}_{\nu}$ the amplitude is sufficiently large to observe drastic difference between linear and nonlinear propagation at $z\sim 1000$. Solitons in Fig. 2 have peak intensities $\sim 2 \cdot 10^{14}~\textrm{W}/\textrm{m}^2$. One can see that after slight decrease of the amplitude at the initial stage both solitons move along the edge without notable modifications, even though they traverse $\sim 100$ $y$-periods of the structure ($\sim 3.2~\textrm{mm}$). Due to helicity of the waveguides, the amplitudes of solitons undergo small periodic oscillations in $z$. To illustrate the importance of focusing (nonlinearity) for localization of the wavepacket, we propagated the same input from $\delta$ branch in the array without nonlinearity, as shown in the middle column in Fig.~\ref{fig:two}. In that case the wavepacket exhibits drastic broadening upon evolution, thus confirming that the states in the first and third columns are indeed topological edge solitons.

\begin{figure*}[t!]
\centering
\includegraphics[width=\linewidth]{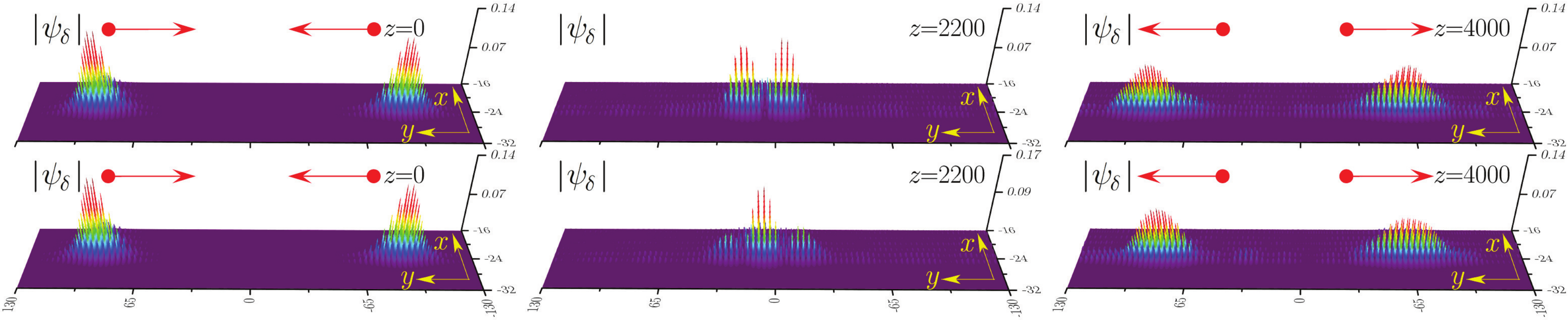}
\caption{Interaction of initially in‐phase (top) and out‐of‐phase (bottom) edge solitons constructed on modes from green branch for $b^{\rm nl}_{\delta}=0.004$. Soliton moving in the positive $y$‐direction corresponds to $k=0.51K$, $b^\prime_\delta=-0.039$, $b^{\prime\prime}_\delta=-0.392$, $\chi_\delta=0.488$, while soliton moving in the negative $y$‐direction corresponds to $k=0.42K$, $b^\prime_\delta=0.028$, $b^{\prime\prime}_\delta=-0.356$, $\chi_\delta=0.377$.}
\label{fig:four}
\end{figure*}

The effective group velocity $-b_{\nu}^\prime$ of the edge states in the dislocated Lieb lattice changes its sign in the Brillouin zone $k\in [0,K]$ for $\beta$ and $\delta$ branches [Fig.~\ref{fig:one}(c)], the phenomenon not encountered in helical honeycomb and non-dislocated Lieb arrays. This implies that solitons bifurcating from the corresponding edge states can propagate in the opposite directions along the $y$ axis or even remain immobile.
To illustrate this, we constructed and propagated solitons bifurcating from  edge states of the $\delta$ branch [Fig.~\ref{fig:one}(b)] with $k$ values corresponding to $b_{\delta}^\prime<0$ (Fig.~\ref{fig:three}, left), $b_{\delta}^\prime=0$ (Fig.~\ref{fig:three}, center), and $b_{\delta}^\prime>0$ (Fig.~\ref{fig:three}, right).

The existence of topological solitons moving in the opposite directions allows one to investigate their collisions. We studied such collisions numerically to confirm exceptional robustness of the solitons. Fig.~\ref{fig:four} shows examples of collision dynamics  of the input pulses  taken the same as in the left and right columns of Fig.~\ref{fig:three}, i.e. the initial separation between them was considerable. Both solitons have nontrivial internal phase distribution because Bloch waves $\phi_{\nu,k}$ on which they are constructed are complex, but to study the impact of the phase on the collision dynamics we imposed additional constant phase difference $\delta\phi$ on input soliton states ($\delta\phi=0$ in the top row and $\delta\phi=\pi$ in the bottom row of Fig.~\ref{fig:four}). Even though nonlinear detuning $b^{\rm nl}_\nu$ is the same for both solitons, the propagation constants $b_{\nu,k}$ for linear states on which solitons are constructed are different, hence interaction scenario depends on the phase difference accumulated at the moment of collision. Apparently, in the top row of Fig.~\ref{fig:four} solitons arrive to collision point approximately out-of-phase that causes destructive interference in collision point, while in the bottom row they collide in-phase that leads to constructive interference. 

Irrespective of the phase difference, solitons pass through each other without noticeable radiation into the bulk, i.e. their interactions are quasi-elastic. This is explained by the fact that the velocities of the solitons are determined by the opposite group velocities of the carrier waves, i.e. by the first-order effect as compared to nonlinear interaction determined by the cross-phase modulation $\sim \langle (|\phi_{\delta,k_1}|^2,|\phi_{\delta,k_2}|^2)\rangle_Z|A_{\delta,k_1}|^2|A_{\delta,k_2}|^2$ (here carrier waves  $\phi_{\delta,k_1}$ and $\phi_{\delta,k_2}$ propagate with opposite group velocities). In other words, irrespective of their phase difference, topological envelope solitons cannot repel each other. This is also seen from corresponding snapshots because two solitons have slightly different penetration depths into the array due to different transverse distributions of carrying Bloch waves. 

\begin{figure}[t!]
\centering
\includegraphics[width=\linewidth]{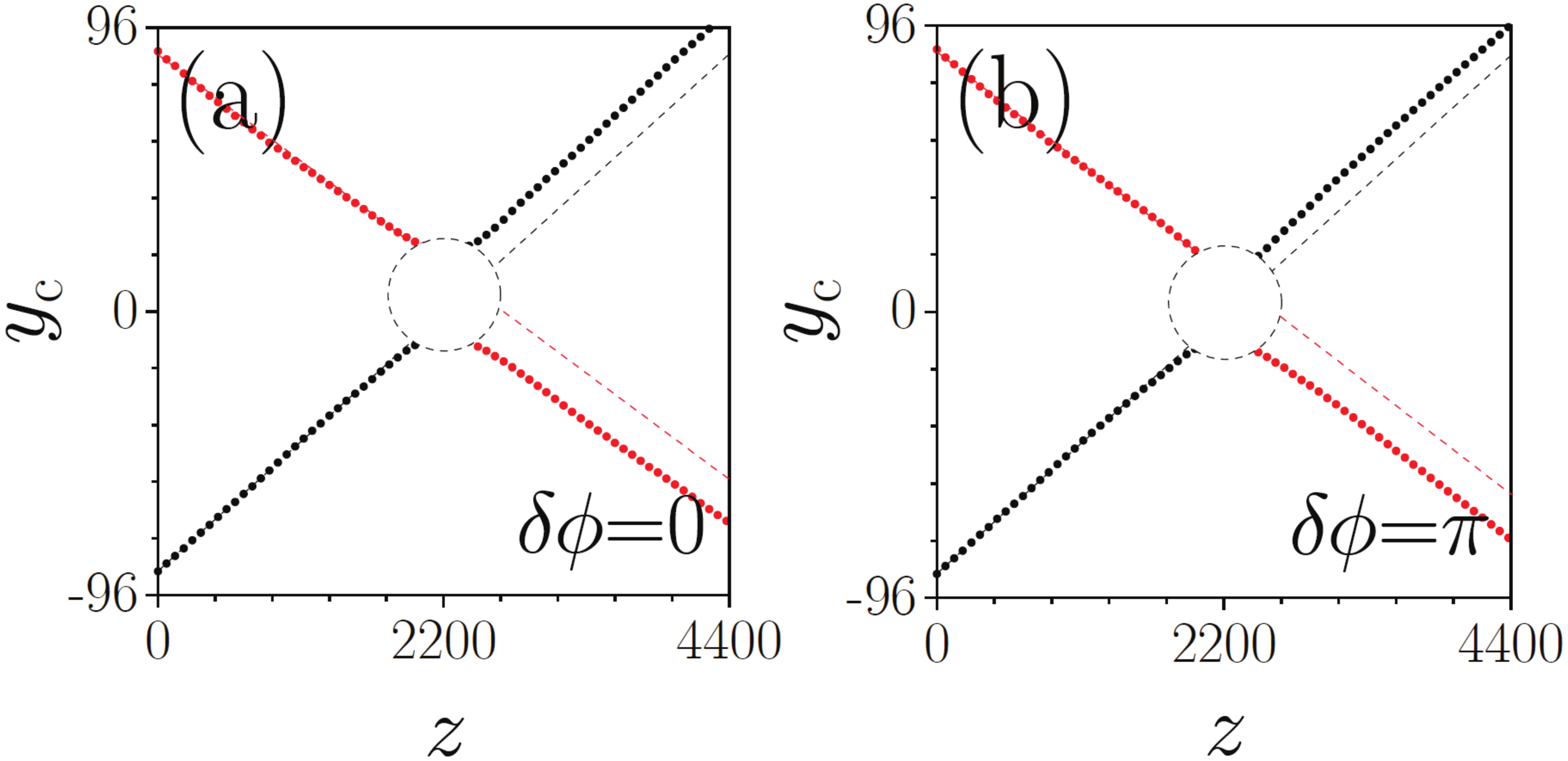}
\caption{Soliton center trajectories corresponding to Fig.~\ref{fig:four}. Dashed lines highlight soliton center shift.}
\label{fig:five}
\end{figure}

Quasi-elasticity of the interaction of topological edge solitons is also seen from the comparison of the soliton trajectories before and after the interactions. The coordinates of soliton centers $y_c$ versus $z$ are shown in Fig.~\ref{fig:five} for two initial phase differences $\delta\phi$. The trajectory of soliton propagating in the positive $y$-direction is shown in black, while the soliton trajectory propagating in the negative $y$-direction is shown in red. The white circle represents the region in which the coordinates of soliton centers are not well-distinguishable. The ``size'' of the interaction region can be estimated as $\ell=\ell_{\delta,k_1}+\ell_{\delta,k_2}$ along the $y$-direction and as $z_0=\ell/(|v_{\delta,k_1}|+|v_{\delta,k_2}|)$ along the $z$-direction. For the chosen parameters (in the dimensionless units) $\ell\approx 19$ and $z_0\approx 300$. As one can see from the comparison of the trajectories before and after the collision (for convenience, in Fig.~\ref{fig:five} we show dashed lines indicating the propagation of the same solitons without interaction), the interaction causes appreciable shift that slightly and non-monotonously changes with phase difference $\delta\phi$.
 
Summarizing, we have shown that a dislocated photonic Lieb arrays support stable edge solitons. In this system edge solitons can propagate in the opposite directions at a given edge of the topological insulator. They are robust with respect to collisions with solitons from the same or from different topological gaps.

\medskip

The authors acknowledge funding from the German research foundation (grant SZ 276/19-1),  RFBR (grant 18-502-12080), and Portuguese Foundation for Science and Technology (FCT) under Contract no. UIDB/00618/2020.

\medskip

\noindent\textbf{Disclosures.} The authors declare no conflicts of interest.

\bibliography{sample}

\end{document}